\documentstyle[12pt]{article}
\catcode`\@=11
\def\makepreprititle{\par
  \begingroup
  \def\thefootnote{\fnsymbol{footnote}}
  \def\
@makefnmark{\hbox
  to 0pt{$^{\@thefnmark}$\hss}}
  \if@twocolumn
  \twocolumn[\@makepreprititle]
  \else \newpage
  \global\@topnum\z@
  \@makepreprititle \fi\thispagestyle{empty}\@thanks
  \endgroup
  \setcounter{footnote}{0}
  \let\makepreprititle\relax
  \let\@makepreprititle\relax
  \gdef\@thanks{}\gdef\@author{}\gdef\@title{}
  \gdef\@preprintnumber{}\gdef\@preprintdate{}\gdef\subtitle{}
  \let\thanks\relax}
\def\preprintnumber#1{\gdef\@preprintnumber{#1}}
\def\preprintdate#1{\gdef\@preprintdate{#1}}
\def\subtitle#1{\gdef\@subtitle{#1}}
\def\@makepreprititle{\newpage
{\def\baselinestretch{1}
  \begin{flushright} \@preprintnumber \par
  \@preprintdate \end{flushright} } \par
  \begin{center}
\vskip 1.5em
  {\LARGE \@title \par} \vskip 2.5em
  {\large \lineskip .5em
  \begin{tabular}[t]{c}\@author
  \end{tabular}\par}
  \vskip 1em {\large \@date} \end{center}
  \par
  \vfil}
\preprintdate{~}
\preprintnumber{~}
\subtitle{}
\def\abstract{\if@twocolumn
\section*{Abstract}
\else \normalsize
\begin{center}
{\bf Abstract\vspace{-.5em}\vspace{0pt}}
\end{center}
\quotation
\addtocounter{page}{-1}
\fi}
\def\endabstract{\if@twocolumn\else\endquotation\fi}
\catcode`\@=12
\title{Heavy Polonyi Field as a Solution of the Polonyi Problem}
\author{Izumi Joichi and Masahiro Yamaguchi}
\date{Department of Physics,  Tohoku University \\
      Sendai 980-77, Japan}
\preprintnumber{TU-465}
\preprintdate{September 1994}
\begin{document}
\makepreprititle
\begin{abstract}
We study properties of the Polonyi field in the renormalizable hidden
sector in supergravity models.  It is shown that radiative corrections
induce a sizable coupling of the Polonyi field to gravitinos as well
as raise its mass to an intermediate scale.  We find it serves a
solution of the Polonyi problem and show that the gravitino abundance
produced by the Polonyi decay is not large.  It is also pointed out
that soft supersymmetry breaking terms induced by this hidden sector
are very restricted.
\end{abstract}

\newcommand{\gsim}{ \mathop{}_{\textstyle \sim}^{\textstyle >} }
\newcommand{\lsim}{ \mathop{}_{\textstyle \sim}^{\textstyle <} }

\newpage

 Hidden sector supersymmetry (SUSY) breaking in supergravity models is
quite a popular scenario which can provide soft-breaking terms in an
effective theory\cite{SUSY}.  The basic idea is that the local SUSY is
spontaneously broken in the hidden sector, which is assumed to couple
with the observable sector only by the gravitational strength.  This
SUSY breaking is transferred to the observable sector through this
very weak interaction.  The scale of the soft-breaking masses becomes
$10^2$ GeV if the SUSY breaking scale in the hidden sector is
$O(10^{10})$ GeV.

However, this phenomenologically successful scenario may cause a
cosmological problem which is referred to as the Polonyi problem
\cite{Polonyi-problem}.  It  was pointed out in the model invented
by Polonyi \cite{Polonyi,Cremmeretal}, whose superpotential is simply
\begin{equation}
   W= \mu^2 (Z+\beta). \label{Polonyi-superpotential}
\end{equation}
Here $Z$ is a chiral multiplet in the hidden sector, $\mu \sim
10^{10}$ GeV and $\beta$ is some constant of order the Planck mass.
The Polonyi field $z$, the scalar component of the $Z$ which is
responsible for the SUSY breaking, has a mass of the gravitino mass
scale with a vacuum expectation value of order the Planck scale.  Its
coherent oscillation, with its initial amplitude of order the Planck
scale, dominates the energy density of the universe as the temperature
goes down.  Its decay, if occurred during or after the primordial
nucleosynthesis, upsets the standard scenario of the big-bang
nucleosynthesis.  Even if it decays before the era of the
nucleosynthesis, the decay produces a huge entropy, diluting the
primordial baryon number asymmetry.  It has recently been
recognized\cite{BKN,CCQR} that many other hidden sector models share
the same cosmological disaster associated with a light, weakly
interacting scalar field.

A solution of the Polonyi problem was already proposed soon after it
was pointed out\cite{DFN,CHRR}.  There the hidden sector is extended
to the following O'Raifeartaigh type model, with
\begin{equation}
   W=\lambda Z(X^2-\mu^2)+m XY. \label{o'raifeataigh}
\end{equation}
Here $X,Y,Z$ are chiral multiplets, $\mu, m$ are mass parameters of
the intermediate energy scale $\sim 10^{10}$ GeV, and $\lambda$ is a
coupling constant.  The Polonyi field $z$ in this case can acquire a
mass of an intermediate scale by one-loop radiative corrections.  At
the same time, $z$ has a sizable coupling to goldstinos, the fermionic
partners of $z$ which are eaten by the gravitino in the superhiggs
mechanism.  The Polonyi field will decay much faster than the case of
the original Polonyi model (\ref{Polonyi-superpotential}) and indeed
there is no entropy production at its decay.  A characteristics of the
O'Raifeartaigh model is that the physics which plays an essential role
to break down supersymmetry is the intermediate physics of order
$10^{10}$ GeV not the Planck scale physics.  The role of the Planck
scale is to mediate the breaking of supersymmetry in the hidden sector
to the observable sector.  In Ref.~\cite{BKN} such a class of models
was termed the renormalizable hidden sector.

In this paper we will investigate properties of the Polonyi field in
the renormalizable hidden sector. We will show that radiative
corrections do not only raise the Polonyi mass generally up to the
intermediate mass scale but also enhance its coupling to the
goldstinos, or the longitudinal modes of the gravitino.  Therefore the
Polonyi field decays before its energy density dominates in the
universe and there is no entropy production due to its decay.
However, if this field dominantly decays to the gravitinos, we have to
examine whether the produced gravitinos may cause cosmological
troubles.  This situation reminds us of the famous gravitino
problems\cite{gravitino1,gravitino2,gravitino3}.  Most severe part of
them is that if an amount of the gravitinos decay radiatively during
or after the primordial nucleosynthesis, the success of the standard
big-bang nucleosynthesis will be destroyed.  To avoid this, we should
lower the reheating temperature of the inflation to reduce the number
density of the gravitinos produced by the scattering processes in the
thermal bath\cite{gravitino3}.  We will calculate the gravitino
abundance produced by the Polonyi decay in the inflationary universe.
We will show that the gravitino number density produced by the Polonyi
decay is smaller than that from the ordinary scattering processes and
thus we do not obtain any further restriction on the reheating
temperature.  Also we will study the structure of the soft terms given
at the flat limit.

Raising up the Polonyi mass in the
renormalizable hidden sector can be understood in the context of
global supersymmetry.   Let us first review the example in
Ref.~\cite{DFN}.  At
the tree-level with the minimal K\"ahler potential, the mass of the
Polonyi field $z$ vanishes.  At the one-loop, terms such as
\begin{equation}
   \frac{h}{M_I^2} \int d^4 x d^4 \theta (\bar Z Z)^2
\end{equation}
are generated in the K\"ahler potential of the effective action.  Here
$\bar Z$ is the hermitian conjugate of the chiral multiplet $Z$, $M_I$
is the intermediate mass scale of order $10^{10}-10^{11}$ GeV and $h$
represents some factor coming from the loop-integral.    This
term gives a mass term of the Polonyi potential
\begin{equation}
  \frac{h}{M_I^2} |\bar F^{\bar z}|^2 \bar z z \sim hM_I^2 \bar z z,
\end{equation}
as well as the interaction of the Polonyi field with the Goldstone
fermions (goldstinos) $\psi^z$
\begin{equation}
   \frac{h}{M_I^2} \bar F^{\bar z} \bar z \psi^z \psi^z
    \sim
     h \bar z \psi^z \psi^z.
\end{equation}

Let us now extend this result to a more general case.  To this end, we
take an effective action approach, where effects of radiative
corrections can be incorporated into  modification of the K\"ahler
potential to a non-minimal one.  We will see that non-renormalizable
terms in the non-minimal K\"ahler potential plays an important role in
our discussion.  Relevant terms of the effective action describing the
hidden sector are thus
\begin{equation}
   \int d^4x d^4\theta K(\Phi, \bar \Phi) +\int d^4x d^2 \theta W(\Phi)
          +h.c.
\label{lagrangian}
\end{equation}
$K$ and $W$ are the K\"ahler potential and the superpotential,
respectively\cite{WessBagger}.  $\Phi^i$ ($\bar \Phi ^{\bar i}$) is a
(anti-)chiral multiplet with components ($\phi^i$, $\psi^i$, $F^i$)
(($\bar
\phi^{\bar i}$, $\bar \psi^{\bar i}$, $\bar F^{\bar i}$)).  For
simplicity,  we do not consider gauge interactions.
  We can extract the scalar
potential from Eq.~(\ref{lagrangian})
\begin{eqnarray}
   V &=& -K_{i \bar j} F^i \bar F^{\bar j} -W_i F^i
         -\bar W_{\bar i} \bar F^{\bar i}
\nonumber \\
   &= & K_{i \bar j} F^i \bar F^{\bar j},
\label{scalar-potential}
\end{eqnarray}
where a subscript $i$ ($\bar i$) means a derivative with respect to a
scalar component $\phi^i$ ($\bar \phi^{\bar i}$).
In the last equality of Eq.~(\ref{scalar-potential}),  we have used
the equations of motion for an auxiliary field $F^i$
and its hermitian conjugate:
\begin{eqnarray}
   F^i &=& -(K^{-1})^{i \bar j} \bar W_{\bar j},
\nonumber \\
   \bar F^{\bar i} & =& -(K^{-1})^{\bar i  j}  W_{ j}.
\end{eqnarray}
Fermion bilinear terms are given as
\begin{equation}
   - \frac{1}{2} \mu_{ij} \psi^i \psi^j,
\end{equation}
with
\begin{equation}
     \mu_{ij}(\phi, \bar \phi) = W_{ij} +K_{ij \bar k} \bar F^{\bar k}.
\end{equation}

Differentiating the scalar potential Eq.~(\ref{scalar-potential}) with
respect to $\phi^i$, we find stationary conditions of the potential
\begin{equation}
   \frac{\partial V}{\partial \phi^i}= -\bar W_{\bar j}
         (\bar F^{\bar j }), _{i} =-\mu_{ij} F^j=0.
\label{stationary-condition}
\end{equation}
To derive the above and some of the subsequent equations, the following
formulae are useful
\begin{eqnarray}
  (F^i), _j \equiv \frac{\partial F^i}{\partial \phi^j}
  &=& -(K^{-1})^{i \bar k} K_{\bar k j l} F^l,
\\
  (\bar F^{\bar i}), _j \equiv
   \frac{\partial \bar F^{\bar i}}{\partial \phi^j}
  &=& -(K^{-1})^{\bar i k} \mu_{k j}.
\end{eqnarray}
{}From Eq.~(\ref{stationary-condition}) it follows that the superfield $Z$
which is responsible for the SUSY breaking, $F^z \neq 0$, has a
vanishing supersymmetric fermion mass $\mu$. This means  the
existence of the goldstino.

   We now consider the scalar mass terms.  The {\em
chirality-conserving}\/ masses are calculated to be
\begin{eqnarray}
M_{i \bar j}^2
&\equiv & \frac{\partial^2 V}{\partial \phi^i \partial \bar \phi^{\bar
j}}
\nonumber \\
 &=& -\mu_{i k} (F^k),_{\bar j} -\mu_{ik, \bar j} F^k
\nonumber \\
&=& \mu_{ik}(K^{-1})^{k \bar l} \bar \mu_{\bar l \bar j}
   -\mu_{ik, \bar j} F^k, \label{mass:global}
\end{eqnarray}
where
\begin{equation}
  \mu_{ij, \bar k}
   \equiv \frac{\partial \mu_{ij}}{\partial \bar \phi^{\bar k}}
   =(K_{ij \bar k \bar l}
    -K_{ij \bar m} (K^{-1})^{\bar m n} K_{n \bar k \bar l})
     \bar F^{\bar l}.
\label{muijk}
\end{equation}
The {\em chirality-flipped}\/ mass terms are
\begin{equation}
  M^2_{ij} \equiv
   \frac{\partial^2 V}{\partial \phi^i \partial \phi^j}
  = -\mu_{ik,j}F^k +\mu_{ik} (K^{-1})^{k \bar l} K_{\bar l m j} F^m.
\end{equation}
In order to get a large mass for the Polonyi field,  the
chirality-conserving mass $M_{z \bar z}^2$ must become large,
because emergence of a large mass only in the chirality-flipped mass
terms would give a negative eigen-value of the mass squared,  implying
the instability of the vacuum.
Since $\mu_{zi}=0$ for any $i$,  the mass of the Polonyi field
$m_z$ is roughly given as
\begin{equation}
    m_z^2 \sim M_{z \bar z}^2=-\mu_{zz, \bar z} F^z.
\end{equation}
In our renormalizable hidden sector,  $K_{i_1 \cdots i_m \bar j_1
\cdots \bar j_n}$ can be as large as $O(M_I^{-m-n+2})$ (times some
factors due to loop-integrals).  Then from
Eq.~(\ref{muijk}), $\mu_{zz, \bar z}$ can be large  as order unity,
in which case the Polonyi field $z$ obtains the intermediate-scale
mass.\footnote{ Our discussion here does not apply to the case  where
$R$-symmetry exists.   For properties of  the renormalizable hidden
sector with the $R$-symmetry,  see Ref.~\cite{BPR}.}

The interactions with the fermions can be seen by taking derivatives of
 $\mu_{ij}$ with respect to $\phi^k$ or $\bar \phi^{\bar k}$.
The Yukawa couplings of the type $\bar \phi^{\bar k} \psi^i \psi^j$ are just
Eq.~(\ref{muijk}) which give the large mass for the Polonyi field.
Therefore when the Polonyi field acquire a large mass by radiative
corrections,  its interaction to the goldstinos becomes inevitably
large.

We will next couple this model to the supergravity.   In terms of the
total K\"ahler potential\cite{Cremmeretal}
\begin{equation}
     G=K+\ln |W|^2,
\end{equation}
the scalar potential is written
\begin{equation}
    V=e^G (G_i (G^{-1})^{i \bar j} G_{\bar j}-3).
\end{equation}
Here we have taken $M=2.4 \times 10^{18}$ GeV to be unity.
Coefficients of the
fermion bilinear terms in the supergravity are
\begin{equation}
\mu_{ij} =e^{G/2} (G_{ij} +G_i G_j-G_k
          (G^{-1})^{k \bar l} G_{\bar l ij}).
\end{equation}
The derivative of $V$ with respect to $\phi^i$ are found to be
\begin{equation}
    V_i=-\mu_{ij} F^j +2m_{3/2}K_{i \bar j}\bar F^{\bar j},
\label{vi:local}
\end{equation}
where
\begin{eqnarray}
    F^i &=& -e^{G/2} (K^{-1})^{i \bar j} G_{\bar j},
\\
   m_{3/2} &=& e^{G/2}.
\end{eqnarray}
After a little algebra,  Eq.~(\ref{vi:local}) can be also written as
\begin{eqnarray}
    V_i &=&-\tilde{\mu}_{ij} F^j,
\nonumber \\
    \tilde \mu_{ij}&=& e^{G/2} \{ G_{ij}
          +(1-\frac{2}{3+e^{-G}V})G_i G_j
         -G_k (K^{-1})^{k \bar l} K_{\bar l ij} \}.
\end{eqnarray}
We can calculate the chirality-conserving masses of the scalars.  The
results are
\begin{eqnarray}
   V_{i \bar j}&=& \mu_{ik}(K^{-1})^{k \bar l} \bar \mu _{\bar l \bar j}
      -F^k \{ K_{i \bar j k \bar l}
         -K_{i k \bar m}(K^{-1})^{\bar m n} K_{n \bar j \bar l} \}
       \bar F^{\bar l}
\nonumber \\
   & & +(V+e^G) K_{i \bar j}-e^G G_i G_{\bar j}. \label{mass:local}
\end{eqnarray}
The last line is of order $m_{3/2}^2$.  Up to these terms,  we can
find a close similarity with the mass terms of the global SUSY case
(\ref{mass:global}).

The coupling of $\psi^i \psi^j \bar \phi ^{\bar k}$ can be easily
obtained, as in the case of the global SUSY,  by differentiating
$\mu_{ij}$ with respect to $\bar \phi^{\bar k}$.  We find it is
\begin{eqnarray}
    \mu_{ij, \bar k}&=&
   \{ K_{ij \bar k \bar l}- K_{ij \bar m}(K^{-1})^{\bar m n}
    K_{n \bar k \bar l} \} \bar F^{\bar l}
\nonumber \\
  & & +G_{\bar k} \mu_{ij} /2 +e^{G/2}(K_{i \bar k}G_j +K_{j \bar
k}G_i)
\nonumber \\
  &=& \{ K_{ij \bar k \bar l}- K_{ij \bar m}(K^{-1})^{\bar m n}
    K_{n \bar k \bar l} \} \bar F^{\bar l} +O(m_{3/2}/M).
\label{coupling:local}
\end{eqnarray}
Substituting the above equation into Eq.~(\ref{mass:local}),
we find
\begin{equation}
    V_{i \bar j}=\mu_{ik} (K^{-1})^{k \bar l} \bar \mu_{\bar l \bar j}
            -\mu_{ik, \bar j} F^{k} +O(m_{3/2}^2),
\end{equation}
which is  the same expression as Eq.~(\ref{mass:global}) for
the global SUSY up to terms of order $m_{3/2}^2$.  Therefore the
arguments we have given in the global SUSY case are still valid after
the model couples to the supergravity.

We are now in a position to discuss the fate of the heavy Polonyi
field of the renormalizable hidden sector in the inflationary
universe.  At the time $t \simeq m_z^{-1}$, the Polonyi field starts coherent
oscillation.  We assume it occurs during the coherent oscillation
 of an inflaton.  The initial energy density of the Polonyi
field oscillation is given
\begin{equation}
   \rho_{z,0} = m_z^2 (\Delta z_0)^2,
\end{equation}
where $\Delta z_0$ stands for the initial amplitude of the
oscillation, which is  typically of order $10^{10}-10^{11}$ GeV
\cite{DFN,CHRR}.  The precise
value of $\Delta z_0$ should depend on  details of the inflation
model.   We consider, for the time being, the case where the inflaton
decays before the Polonyi decay.
The energy density of the Polonyi field evolves like that of a
non-relativistic particle,  and hence at the decay of the
inflaton (and the reheating of the universe) it becomes
\begin{equation}
   \rho_z (t=\Gamma_\varphi^{-1})
  \simeq \rho_{z,0}\left( \frac{\Gamma_\varphi}{m_z} \right)^2
  = \Gamma_\varphi^2 (\Delta z_0)^2,
\end{equation}
where $\Gamma_\varphi$ denotes the decay width of the inflaton.
Note that this decay width is related to the reheating temperature $T_{RH}$ of
the inflationary universe as
\begin{equation}
 \Gamma_\varphi \simeq 2 H=0.66 g_*^{1/2}(T_{RH}) \frac{T_{RH}^2}{M}
\label{reheating-temp}
\end{equation}
with  $g_*$ being the effective degree of freedom of the radiation.
(We follow the notation of Ref.~\cite{KolbTurner}.)
After the reheating,  the universe becomes radiation dominant.  The
ratio of the energy density of the Polonyi field to the entropy
density $\rho_z / s$ remains constant until it decays.
Its decay width is roughly
\begin{equation}
  \Gamma_z \sim \frac{m_z^5}{m_{3/2}^2 M^2}.
\end{equation}
Here we assume that it dominantly decays to the gravitinos.  If
another mode dominates,  the decay width becomes even larger.
  We can see that the Polonyi field decays before its energy
density dominates the universe, as far as
\begin{equation}
   \frac{\Gamma_z}{\Gamma_\varphi} \gsim \left( \frac{\Delta z_0}{M}
\right)^4,
\end{equation}
which is satisfied for our scenario with $m_z, \Delta z_0 \sim
10^{10}$ GeV.  Hence there is no entropy crisis  in the
renormalizable-hidden-sector SUSY breaking.

However,  if the Polonyi field decays to the gravitinos dominantly,
this decay may cause another cosmological problem.   Let us calculate
how much the gravitinos are produced by this decay.  Assuming that
it decays to two gravitinos with $100\%$ branching ratio,  we find the
number density of the gravitinos just after the decay of the Polonyi
field is
\begin{equation}
      n_{3/2}=2n_z=2 \frac{\rho_z}{m_z}.
\end{equation}
The gravitino, after produced, does not interact with each other and
hence the gravitino number density-to-entropy ratio $n_{3/2}/s$ remains
constant.
We therefore find that
\begin{eqnarray}
\frac{n_{3/2}}{s} &=& 2 \frac{\rho_z}{m_z s}(t=\Gamma_z^{-1})
\nonumber \\
  & =& 2 \frac{\rho_z}{m_z s}(t=\Gamma_\varphi^{-1})
\nonumber \\
  &\simeq &2 \frac{\Gamma_\varphi^2 (\Delta z_0)^2}{m_z s}
(t=\Gamma_\varphi^{-1}).
\end{eqnarray}
Substituting
\begin{equation}
  s=0.44 g_{* s} T^3
\end{equation}
and Eq.~(\ref{reheating-temp}) into the above and using $g_{* s}=g_*$,
we find
\begin{equation}
\frac{n_{3/2}}{s}\simeq 2.0 \frac{T_{RH}}{m_z} \left(\frac{\Delta z_0}{M}
\right)^2.
\label{n3/2/s}
\end{equation}

Since the gravitino does not interact even  with the thermal bath, its
momentum,
which is large ($\sim m_z$) when produced, get just red-shifted.
We can see at the decay of the gravitino,  it is non-relativistic and its
energy is just the same as  its mass.  Therefore the energy density of
the gravitinos $\rho_{3/2}$ at the decay satisfies
\begin{equation}
  \frac{\rho_{3/2}}{s}
=m_{3/2} \frac{n_{3/2}}{s}
\simeq 2.0 \frac{m_{3/2}}{m_z} \left( \frac{\Delta z_0}{M} \right)^2
T_{RH}.
\label{rho3/2/s}
\end{equation}
Note that  a small factor $m_{3/2}/m_z $ is missed in Ref.~\cite{CHRR} and our
result is consistent with that of Ref.~\cite{BPR}.

Gravitinos are also produced through
scattering processes in the thermal bath.
When the gravitino weighs less than 1 TeV, it decays during or after
the period of the primordial nucleosynthesis.  In order that
the decay should not upset the success of the standard big-bang
nucleosynthesis,  the energies released by the gravitino decays should
be small enough.  This gives the upperbound of the reheating
temperature.

To see whether the gravitino production from the Polonyi decay will
affect this standard scenario, we want to compare Eq.~(\ref{n3/2/s})
with the gravitino number-to-entropy ratio coming from the scattering
processes.  Here we use the recent result of Kawasaki and Moroi
\cite{KawasakiMoroi}.  They calculated the production cross section of
the gravitinos in the minimal SUSY standard model (MSSM) to obtain the
ratio $n_{3/2}/n_{rad}$, where $n_{rad}$ is the number density of a
relativistic real scalar boson.  The entropy density $s$ is related to
$n_{rad}$ as $s=3.60 g_{*s} n_{rad}$ with $g_{*s}(T \ll 1 {\rm MeV})
=3.36$.  From Eq.~(7) of Ref.~\cite{KawasakiMoroi}, we therefore find
that the scattering processes give approximately
\begin{equation}
\frac{n_{3/2}}{s}  |_{scatterings}
= 1.8 \times 10^{-12} \left( \frac{T_{RH}}{10^{10} {\rm GeV}} \right)
   \{ 1- 0.0232 \log (T_{RH}/10^{10} {\rm GeV} ) \}.
\end{equation}
On the other hand,  the same ratio coming from the Polonyi decay
(\ref{n3/2/s}) can be rewritten
\begin{equation}
 \frac{n_{3/2}}{s} |_{Polonyi \ decay} \simeq
   3.4 \times 10^{-15} \left( \frac{T_{RH}}{10^{10} {\rm GeV}} \right)
              \left( \frac{10^{10} {\rm GeV}}{m_z} \right)
           \left( \frac{\Delta z_0}{10^{11} {\rm GeV}} \right)^2.
\end{equation}
For $m_z \sim 10^{10} $ GeV and $\Delta z_0 \sim 10^{11}$ GeV, we find
the number of the gravitinos produced by the decay of the Polonyi
field is by three orders of magnitude smaller than that by the
scattering processes.  Thus unless the Polonyi mass is smaller than
$\sim 10^7$ GeV,  we do not obtain any new constraint on the
reheating temperature in this scenario.
Note that when the reheating temperature is low,  the Polonyi field
may decay during the coherent oscillation of the inflaton.  However, a
similar analysis shows that the gravitino-to-entropy ratio in this
case is the same as Eq.~(\ref{n3/2/s}).

Finally we would like to discuss the coupling of the hidden sector to
the observable sector.  In our renormalizable hidden sector models,
non-renormalizable terms in the K\"ahler potential between the
observable sector and the hidden sector should only appear as
radiative corrections.  Thus we can, as an approximation, take the
minimal K\"ahler potential for the observable sector.  Namely we have
\begin{eqnarray}
    K &=& \hat K(z, \bar z) + \sum_i |y^i|^2,\\
    W &=& h(z) +g(y).
\end{eqnarray}
Here $z$ ($y$) is a field in the hidden (observable) sector and $\hat
K$ admits non-minimality as in the previous arguments.  Recalling that
$\hat K_z =O(z) \ll M$, we obtain the scalar potential at the flat
limit
\begin{equation}
V_{flat~limit}=\sum_i |\hat g_i|^2 +(|m_{3/2}|^2 +\langle V \rangle /M^2)
   \sum_i |y^i|^2 +m_{3/2}^* (\sum_i y^i \hat g_i -3 \hat g) +h.c.,
\end{equation}
where $\hat g=e^{\hat K/2} g$ is the (normalized) superpotential and a
phase   of the gravitino mass is taken into account.
With the vanishing cosmological constant $\langle V \rangle =0$,  we
find soft-terms in the scalar sector can be specified only  by the
gravitino mass.  Indeed we have
\begin{itemize}
\item universal scalar mass:
\begin{equation}
                    m_0=|m_{3/2}|,
\end{equation}
\item trilinear scalar coupling:
\begin{equation}
                       A=0,
\end{equation}
\item Higgs mixing mass:
\begin{equation}
                    B=-m_{3/2}^*.
\end{equation}
\end{itemize}
It is an interesting task to study low-energy consequences of these
specific soft-terms.
Note that we can obtain a non-vanishing gaugino mass $M_{1/2}$ by
taking a non-canonical gauge kinetic function.

\

\

We thank T. Yanagida for helpful discussions.

\end{document}